\documentclass[11pt]{article}
\usepackage[T1]{fontenc}
\usepackage{fullpage}
\usepackage{subcaption}
\usepackage{graphicx}
\usepackage{amsmath,amssymb}
\usepackage{algorithmic}
\usepackage{hyperref}

\begin{document}

\title{\rule{\linewidth}{2pt} \\ [0.4em]
Adaptive Training for Nautical Rules of the Road \\ [0.0em]
\rule{\linewidth}{2pt}
}

\author{Amit Dutta \\
\small Evolutionary Computing Systems Lab (ECSL) \\
\small Dept. of Computer Science and Engineering \\
\small University of Nevada, Reno \\
\small amitd@unr.edu
\and
Sushil J. Louis \\
\small Evolutionary Computing Systems Lab (ECSL) \\
\small Dept. of Computer Science and Engineering \\
\small University of Nevada, Reno \\
\small sushil@unr.edu \\
}
\date{January 25, 2026}
\maketitle              

\begin{abstract}
Knowledge of the nautical rules of the road is essential for safe ship navigation and collision avoidance. We evaluated adaptive and non-adaptive versions of a ship-driving simulation trainer designed to assess and improve students' knowledge and application of these rules. We randomly assigned 30 university students to an adaptive or non-adaptive training condition and measured learning using pretest and post-test scores. Students who received adaptive training achieved significantly higher post-test scores than those who received non-adaptive training ($p < 0.0001$). After the post-test, all students experienced both versions of the trainer and compared them in a survey. Of the 30 students, $73\%$ judged the adaptive trainer more effective, and 22 rated it "very engaging," compared with 9 who gave the non-adaptive trainer the same rating. These findings provide evidence that adapting scenario difficulty and providing immediate, context-sensitive feedback can improve both learning outcomes and student engagement in simulation-based training.


\end{abstract}
\section{Introduction}

Safe vessel navigation requires mariners to understand and correctly apply the International Regulations for Preventing Collisions at Sea (COLREGs)~\cite{imo1972convention}. Navigation errors and COLREGs violations contribute to ship collisions that can cause injuries, fatalities, property damage, and environmental harm~\cite{maternova2023human, ugurlu2022analysis}. Although the COLREGs provide a common framework for preventing collisions, students often find them difficult to master, particularly during the early stages of maritime education~\cite{demirel2015further}.

Real-world navigation training presents additional challenges. Students need opportunities to recognize collision risks, interpret vessel movements, apply the appropriate rules, and select safe courses of action. Practicing these skills aboard an actual vessel, however, can be costly and potentially dangerous. Simulation-based training provides a controlled environment in which students can practice decision making, make mistakes, and receive feedback without creating real-world risks. Previous research has demonstrated the value of simulation-based training in safety-critical domains such as aviation~\cite{hays1992flight}.

Repeated practice alone does not guarantee efficient learning. Training effectiveness also depends on how a system selects exercises, adjusts their difficulty, and provides feedback~\cite{landsberg2012review,Nabizadeh2020LearningPP,song2024implementing}. Traditional computer-based training systems commonly present the same sequence or level of material to all students. This one-size-fits-all approach may give some students tasks that are too easy while giving others tasks that are too difficult. Both situations can slow learning and reduce engagement~\cite{landsberg2012evaluation}. Adaptive training addresses this limitation by modifying the learning experience in response to each student's performance. An adaptive system can adjust task difficulty, select appropriate content, and provide feedback as the student progresses. A growing body of research suggests that these adaptations can improve learning efficiency and engagement~\cite{debeer2021effect,landsberg2012review,landsberg2012evaluation,metzler2009does}.

\begin{figure}[htbp]
    \centering
        \includegraphics[width=\textwidth]{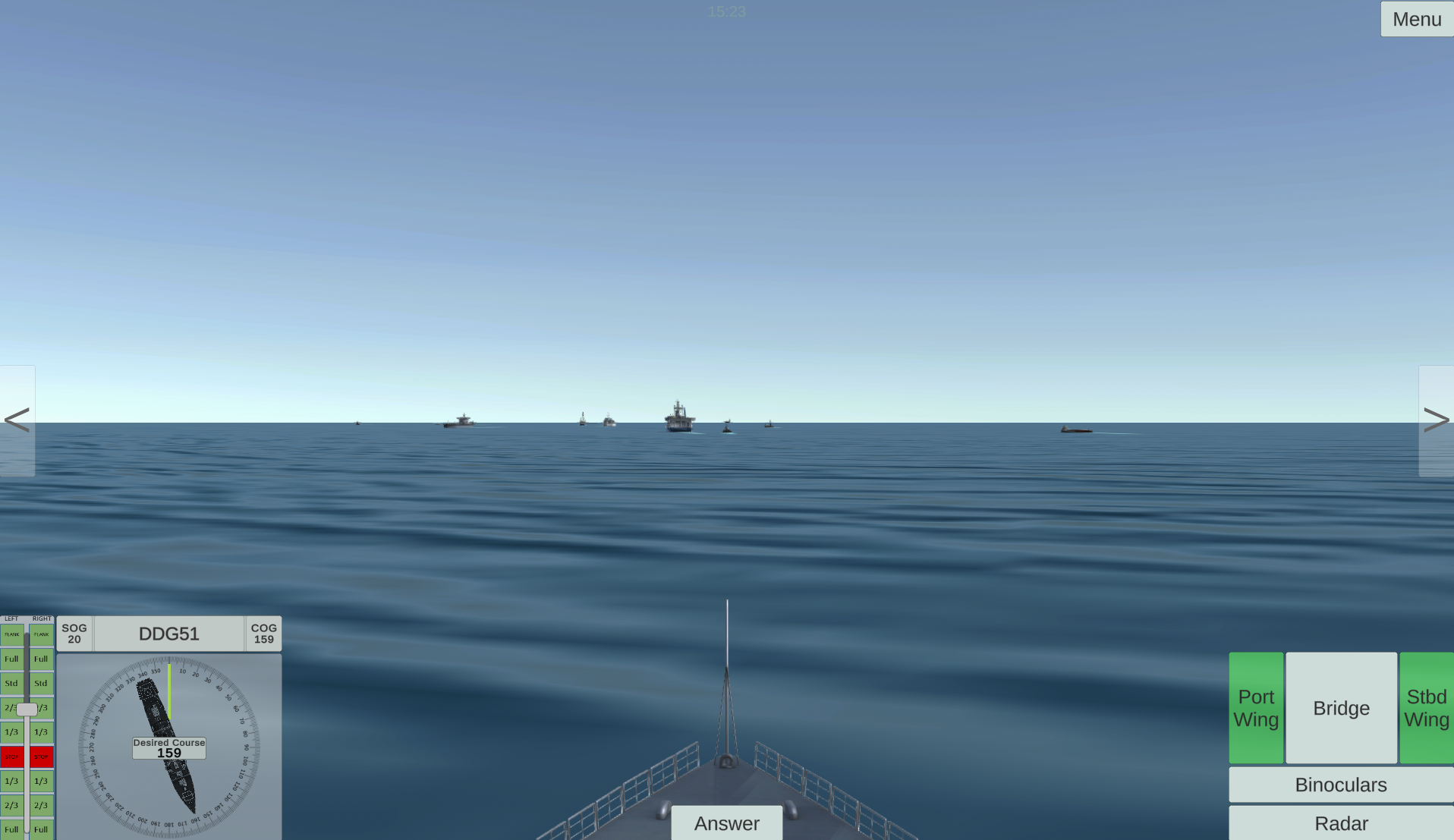}
        \caption{A screenshot of the identical RAFT and RoR interface. The student evaluates the scenario and chooses a course of action in accordance with the nautical rules of the road.}
        \label{RaftScreenshot}
\end{figure}

This paper compares two versions of a simulation trainer for learning and applying the nautical rules of the road. The Rules of the Road Adaptive Fleet Training system (RAFT) provides adaptive training, whereas the Rules of the Road system (RoR) provides non-adaptive training. Both systems present the same instructional content through identical user interfaces, as shown in Figure~\ref{RaftScreenshot}. They also use the same scenario-generation framework. A student observes a simulated navigation scenario, identifies the relevant collision risk, and recommends a course of action that complies with COLREGs.
The two trainers differ in how they control difficulty and deliver feedback. RoR has three fixed difficulty levels: novice, intermediate, and advanced. A student or instructor selects one of these levels, and RoR generates scenarios at the selected level. In contrast, RAFT represents difficulty as a continuous value between $0$ and $1$. After each scenario, RAFT evaluates the student's performance and adjusts this difficulty value before generating the next scenario. The resulting difficulty value determines properties of the next scenario; the time of day, the number of vessels, and the time remaining before the closest point of approach. RAFT therefore presents increasingly challenging scenarios as a student's performance improves. In addition, following adaptive training principles, RAFT provides immediate context-sensitive feedback after each scenario in contrast to RoR which provides feedback only at the end of a quiz. Because the two systems otherwise share the same content, interface, and scenario-generation framework, we can compare adaptive training against non-adaptive training. 

We conducted a user study to compare RAFT against RoR. Students in the RAFT condition achieved significantly higher post-test scores than those in the RoR condition. In addition, after exposure to both trainers, students overwhelmingly preferred RAFT over RoR across several metrics. These results demonstrate the superior engagement and learning outcomes made possible through adaptive training.

This work makes three contributions. First, we describe an adaptive simulation trainer that adjusts the difficulty of automatically generated collision-risk scenarios in response to student performance. Second, we compare adaptive and non-adaptive versions of the trainer while holding their instructional content, scenario framework, and user interface constant. Third, we evaluate the systems using objective learning measures and students' assessments of effectiveness, engagement, and feedback. Together, these results provide evidence that performance-based difficulty adaptation can improve learning and user engagement in a nautical rules-of-the-road simulation.

The rest of this paper is structured as follows. The next section
describes prior work in adaptive training. Section~\ref{Sim} details
the simulation trainer and learning task. We subsequently specify the
design of our experiment. Section~\ref{Results} provides results and the last section summarizes the work, discusses improvements, and provides directions for future work.

\section{Prior Work}

Adaptive training systems tailor instruction to individual learners by modifying task content, difficulty, or feedback in response to learner performance~\cite{brusilovsky2001adaptive,kelley1969adaptive}. Researchers have studied adaptive training across educational, medical, industrial, and military applications, and a recent systematic review summarizes advances in adaptive virtual-reality training~\cite{zahabi2020adaptive}. Although many studies report that adaptive training improves learning or engagement, results vary across tasks, populations, and adaptation methods~\cite{diarchangel2024serious,finseth2021effectiveness,metzler2009does,sharek2015investigating}.

Designing an effective adaptive training system requires identifying appropriate variables to adapt, measuring learner performance, and determining when and how strongly to modify the training experience. Vanbecelaere et al.\ investigated these design decisions in a digital reading game~\cite{vanbecelaere2020effectiveness}. Results showed that children improved their phonological awareness and letter knowledge under all three experimental conditions, but the researchers found no significant differences in cognitive or non-cognitive outcomes between the adaptive and non-adaptive systems. They concluded that adaptive learning systems may require more fine-grained adaptation algorithms. In subsequent work, Debeer et al.\ analyzed learning efficiency in an educational game and found that adaptive training improved efficiency relative to non-adaptive training~\cite{debeer2021effect}. Together, these studies show that adaptation can improve learning, but its effectiveness depends on how the system models performance and adjusts instruction. 

Researchers have also reported mixed results in simulation-based professional training. Pham compared adaptive and non-adaptive laparoscopic surgery trainers and found no significant difference in training effectiveness between the two conditions~\cite{pham2005smart}. However, participants strongly preferred the adaptive version in a post-training survey. This result suggests that adaptation may improve the learner's experience even when measured performance does not improve. Other researchers have applied adaptive techniques to rehabilitation and motor-skill training. Heloir designed a self-adaptive architecture for upper-limb rehabilitation, while Turakhia investigated physical tools that change shape in response to learner performance~\cite{heloir2015design,turakhia2021physical}. These systems illustrate how adaptive methods can support training tasks that involve physical as well as cognitive skills. 

Adaptive training has also produced promising results in safety-critical applications. Schwaninger used adaptive computer-based training to improve the performance of X-ray security screeners and reported that screening performance nearly doubled after training~\cite{schwaninger2007adaptive}. Because that study did not include a non-adaptive comparison condition, it demonstrated improvement after adaptive training but did not isolate the effect of adaptation itself. Finseth evaluated adaptive training for stress inoculation in a simulated astronaut task, further extending adaptive methods to demanding operational environments~\cite{finseth2021effectiveness}.

The work most closely related to RAFT examines adaptive training for maritime decision-making tasks. Landsberg and colleagues evaluated an adaptive system for submarine periscope operations, and DiArchangel and Louis developed a serious game for recognizing target angles~\cite{landsberg2012evaluation,diarchangel2024serious}. These tasks require trainees to interpret the orientation of a target vessel relative to their own submarine or surface vessel. The systems adapted task difficulty and feedback according to trainee performance. RAFT applies a similar performance-based approach to a broader navigational task. Rather than focusing only on target orientation, RAFT requires students to interpret collision-risk scenarios, identify the applicable nautical rule of the road, and recommend an appropriate course of action. After each scenario, RAFT uses the student's score to adjust the difficulty of the next scenario and provides immediate, context-sensitive feedback. We compare RAFT with RoR, a non-adaptive trainer that presents the same content through the same user interface but uses fixed difficulty levels and delayed feedback. This comparison allows us to evaluate whether the adaptive training configuration improves learning outcomes, response times, and students' assessments of the training experience. We next describe RAFT in more detail.

\section{Rules of the Road Adaptive Fleet Training (RAFT)}{\label{Sim}}

The International Regulations for Preventing Collisions at Sea (COLREGs) are global rules set by the International Maritime Organization (IMO) to reduce the risk of vessel collisions in international waters~\cite{imo1972convention}. These regulations dictate how vessels maneuver to reduce risk and avoid collisions. All vessels must comply with these rules to avoid accidents. The COLREGs are regularly updated to keep pace with advancements in maritime navigation.

The Rules of the Road Adaptive Fleet Training (RAFT) system generates realistic collision-risk scenarios and evaluates how well students apply the COLREGs. RAFT currently generates three common encounter types: crossing, head-on, and overtaking. In each scenario, the student operates an ownship and evaluates the surrounding vessels to identify a potential collision. The student can change viewpoints and use the simulated radar display and voyage management system to gather information about nearby vessels.

\begin{figure}[htbp]
    \centering
    \begin{subfigure}{0.49\textwidth}
        \centering
        \includegraphics[width=\textwidth]{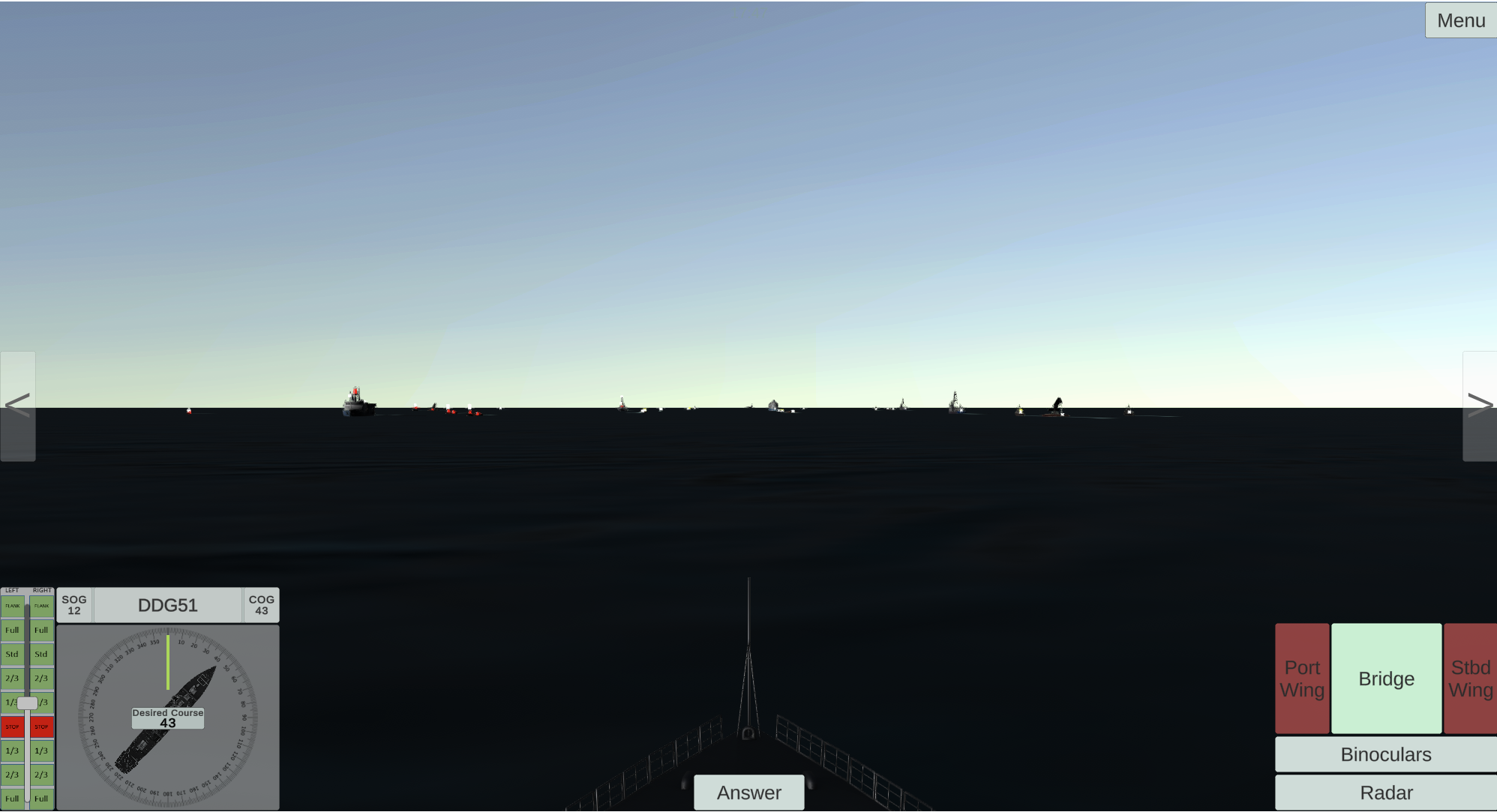}
        \caption{}
        \label{advanceEvening}
    \end{subfigure}%
    \hfill
    \begin{subfigure}{0.49\textwidth}
        \centering
        \includegraphics[width=0.95\textwidth]{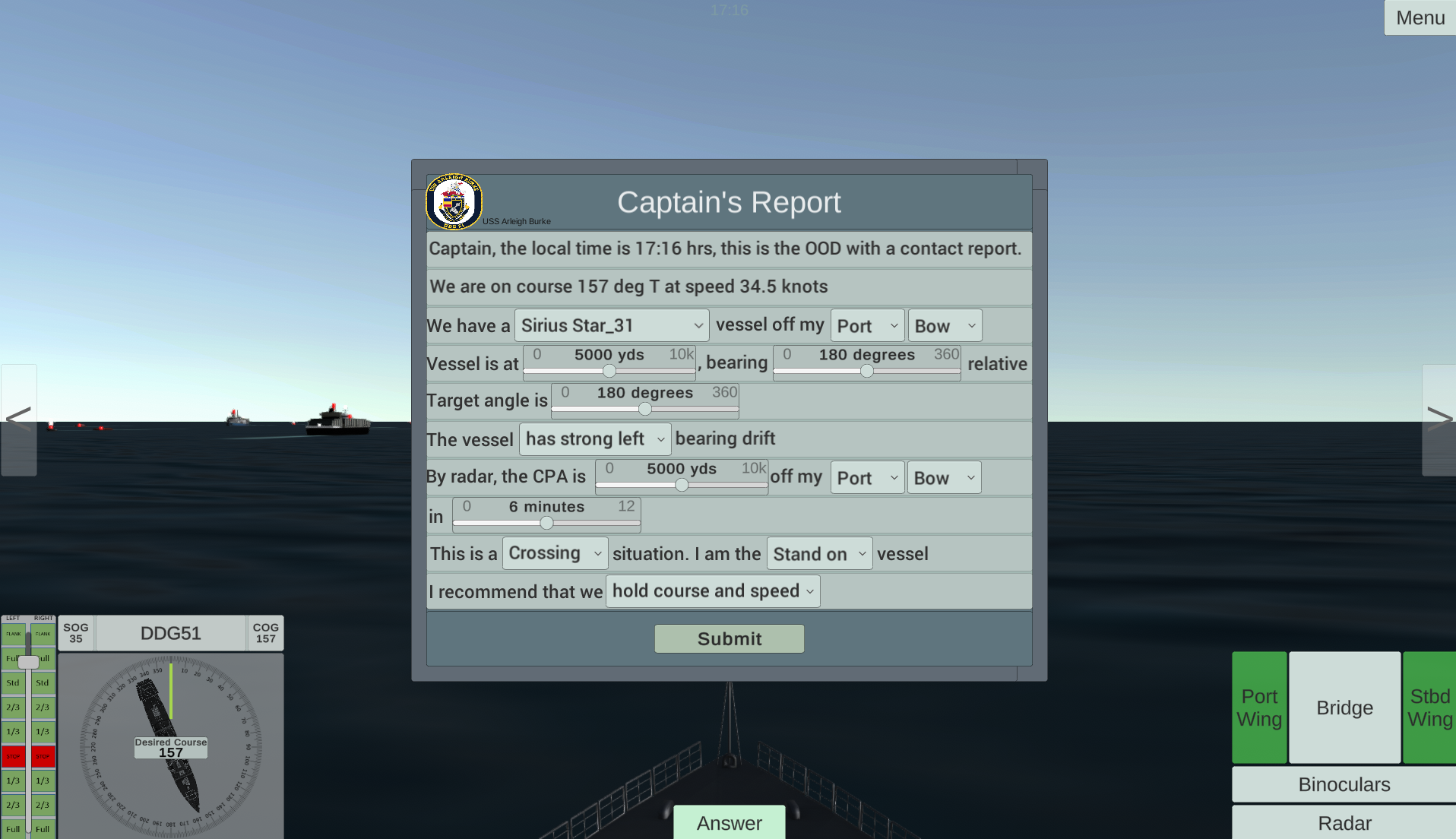}
        \caption{}
        \label{captainsReport}
    \end{subfigure}
    \caption{~\ref{advanceEvening} shows a screenshot from RAFT. The answer button at center bottom brings up the Captain's report (\ref{captainsReport}) that students complete to go to the next scenario. Simulation controls on the bottom right of the interface enable students to change viewpoints and bring up the radar system.}
    \label{fig:raft-scenario-report}
\end{figure}

Figure~\ref{fig:raft-scenario-report} shows the two main components of the training interface. Figure~\ref{advanceEvening} displays a late-evening scenario in which vessel running lights provide important cues about each vessel's position and orientation. The student selects the \emph{Answer} button to open the Captain's Report shown in Figure~\ref{captainsReport}.

\begin{figure}[htbp]
    \centering
        \includegraphics[width=4in]{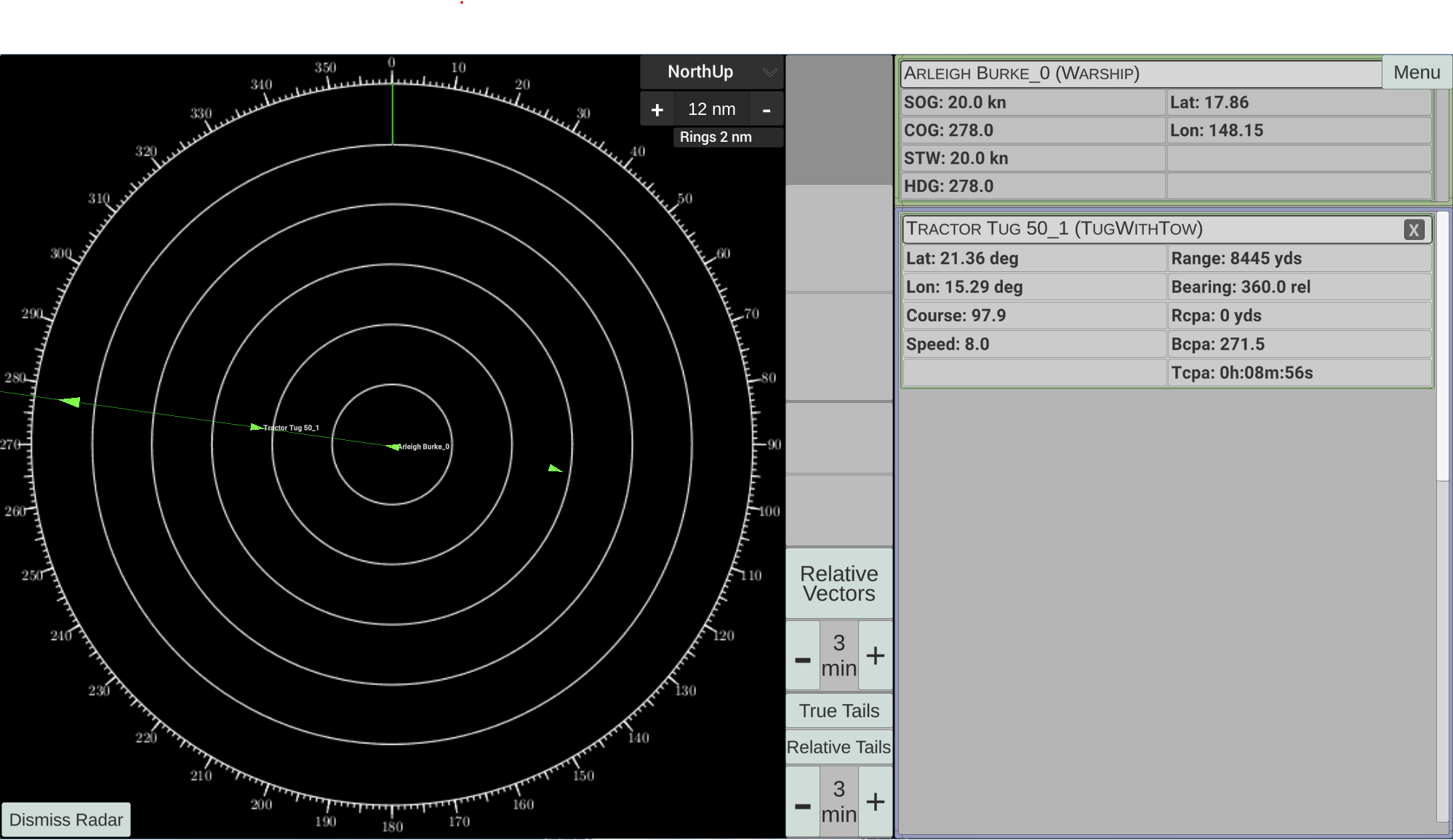}
        \caption{A screenshot of the RAFT/RoR radar interface that provides information on selected vessels.}
        \label{RaftRadarScreenshot}
\end{figure}

The Captain's Report, a navy standardized form filled out by the Officer Of the Deck (OOD) in the real-world, structures the student's analysis of the scenario. The student first identifies the vessel that presents a collision risk, which we refer to as the target vessel. The student then fills the Captain's Report with the target's location, range, relative bearing, bearing drift, and target angle. These values describe the target's position and motion relative to ownship. The student also reports three measures associated with the closest point of approach (CPA): the range at CPA (RCPA), the time to CPA (TCPA), and the resulting collision risk. The student can calculate these values by observation (manually) or obtain relevant information from the radar display shown in Figure~\ref{RaftRadarScreenshot}. Finally, the student identifies the applicable nautical rule of the road and recommends an appropriate course of action.

The Captain's Report reflects the type of information that an officer of the deck communicates, over a radio or ship's phone, to a captain who is not on the bridge. By reporting the relevant vessels, their relative positions and motion, the collision risk, and a recommended action, the OOD provides the captain with the information needed to understand the situation and approve or revise the recommendation.

When a student submits a Captain's Report, RAFT scores each response component. The system uses the resulting scenario score to update its difficulty variable and automatically generate the next scenario at the updated difficulty level. RAFT also gives the student immediate feedback about correct and incorrect responses. The following subsections describe the difficulty model, adaptation procedure, and feedback mechanism.

\subsection{Difficulty}

RAFT represents scenario difficulty with a continuous variable $\delta \in [0,1]$. After a student completes a scenario, RAFT updates $\delta$ from the student's score and uses the new value to configure the next scenario. In contrast, RoR asks the student or instructor to choose one of three fixed difficulty levels and does not change that level in response to student performance. 

RAFT uses $\delta$ to control three scenario variables: 
\begin{itemize} 
    \item \textbf{Time of day (TOD).} Daylight allows students to observe vessel shapes and orientations directly. At night, students must interpret the positions, colors, and visible arcs of vessel running lights. Night scenarios therefore require additional knowledge and present greater difficulty. 
    \item \textbf{Time to closest point of approach (TCPA).} TCPA measures the time remaining before two vessels reach their closest point of approach. A shorter TCPA gives the student less time to evaluate the encounter and choose an action. RAFT therefore associates shorter TCPA values with greater difficulty. 
    \item \textbf{Number of ships in the scenario (NSHIP).} As the number of vessels increases, the student must evaluate more potential collision risks and identify the vessel or vessels that require attention. RAFT therefore associates larger numbers of ships with greater difficulty. 
\end{itemize}
We would therefore like to start new students at a low level of difficulty corresponding to lower $\delta$s. In such cases, the student receives ample time before the closest point of approach, operates in daylight, and encounters relatively few vessels. As students learn to perform better, RAFT increases $\delta$ and generates more challenging scenarios. These scenarios may occur later in the day or at night, include denser traffic, and provide less time before the closest point of approach.  We describe how a student's performance score on a scenario changes the value of $\delta$ and how $\delta$ then determines the values of the three scenario variables above.

\subsection{Performance-Based Difficulty Adaptation}

RAFT updates difficulty after the student submits a Captain's Report with $16$ scored components. For each component $i \in \{1,\ldots,16\}$, RAFT calculates a component score $C_i \in [0,1]$ and the overall scenario score is a weighted sum of component scores as specified in Equation~\ref{eq:scenario-score}.  
\begin{equation} 
    S = \sum_{i=1}^{16} w_i C_i , 
    \label{eq:scenario-score} 
\end{equation}
where $w_i$ denotes the weight assigned to component $i$. The weights were set by our subject matter experts. RAFT normalizes the overall scenario score $S$ to the interval $[0,1]$. 

\begin{figure}[htbp]
    \centering
    \begin{subfigure}[b]{0.33\textwidth}
        \centering
        \includegraphics[width=1.5in]{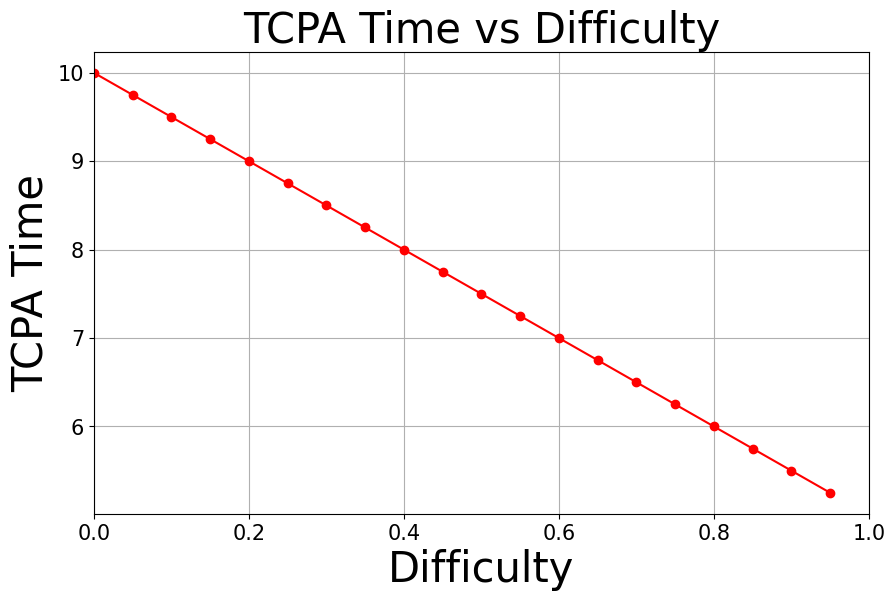}
        \caption{TCPA vs Difficulty}
        \label{fig:diff_cpa}
    \end{subfigure}%
    \begin{subfigure}[b]{0.33\textwidth}
        \centering
        \includegraphics[width=1.5in]{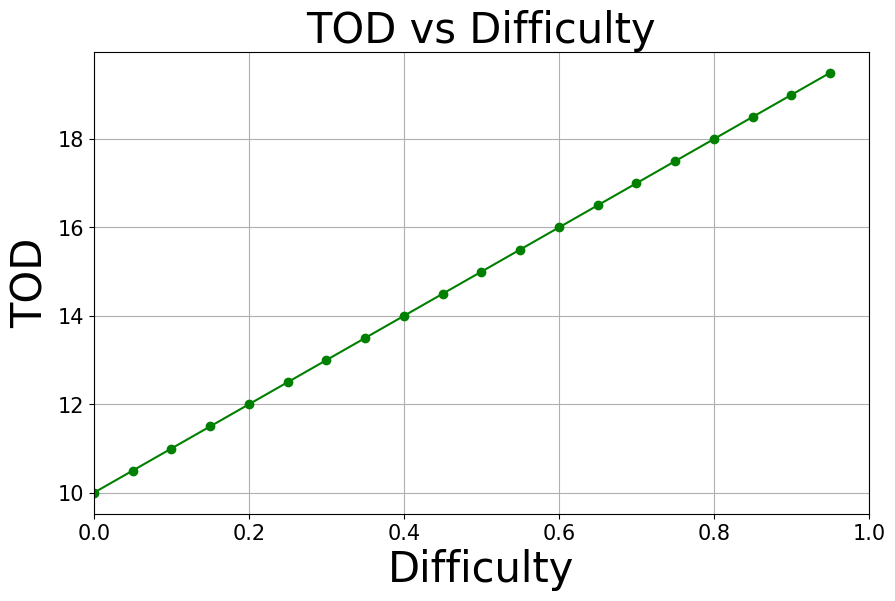}
        \caption{TOD vs Difficulty}
        \label{fig:diff_tod}
    \end{subfigure}%
    \begin{subfigure}[b]{0.33\textwidth}
        \centering
        \includegraphics[width=1.5in]{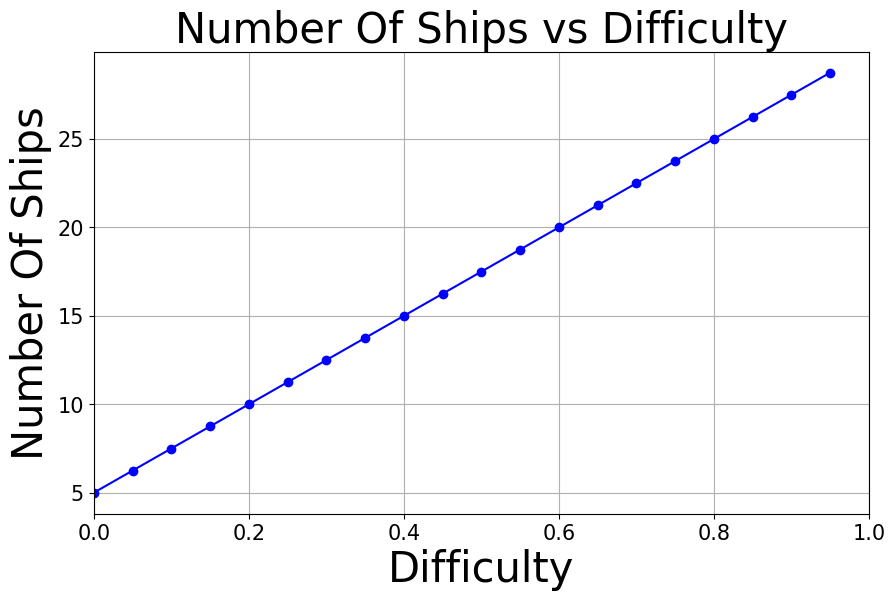}
        \caption{NSHIP vs Difficulty}
        \label{fig:diff_nship}
    \end{subfigure}
    \caption{Plots of difficulty on the horizontal axis versus TCPA, Time of Day (TOD), and number of ships (NSHIP) on the vertical axis. Given a difficulty value, these plots determine the corresponding scenario variable values in RAFT.}
    \label{fig:difficulty_curves}
\end{figure}

Let $t$ index the sequence of scenarios completed by a student. RAFT updates the difficulty of the next scenario according to 
\begin{equation}
    \delta_{t+1} = \delta_t + 0.05 S_t , 
\label{eq:difficulty-update} 
\end{equation} 
where $\delta_t$ is the difficulty of scenario $t$ and $S_t$ is the student's normalized score on that scenario. Under this rule, a higher score produces a larger increase in difficulty. We selected the coefficient $0.05$ after experimenting with several values for the university-student population in this study. Other trainee populations may require a different coefficient. A future system could also adapt this coefficient as the student progresses. After calculating $\delta_{t+1}$, RAFT maps the updated difficulty to values for TCPA, TOD, and NSHIP. Figure~\ref{fig:difficulty_curves} shows these mappings. As difficulty increases, TCPA decreases, while TOD and NSHIP increase.

These linear mappings generate scenario variables from a $\delta_t$ value as follows. Suppose that RAFT calculates a difficulty of $\delta=0.4$. At this difficulty, the mappings in Figure~\ref{fig:difficulty_curves} specify a TCPA of $8$ minutes, a time of day of $1400$ hours, and $15$ ships. RAFT then randomly selects one of the three supported encounter types. If it selects a crossing encounter in which the student's vessel must give way, RAFT generates a 1400-hour scenario containing 15 vessels. One of those vessels follows a crossing course that will reach its closest point of approach with the student's vessel in 8 minutes. We add, that RAFT adds a small amount of randomness ($\pm 10\%$) to the mapped scenario-variable values so that students do not repeatedly encounter identical configurations.

Every student in the adaptive condition in our study begins with $\delta_0 = 0.2$. For comparison with RoR, we divide the continuous RAFT difficulty range into three intervals: 
\begin{equation} 
    \begin{aligned} 
        0.00 \leq \delta \leq 0.33 & \quad \text{corresponds to novice}, \\ 
        0.34 \leq \delta \leq 0.67 & \quad \text{corresponds to intermediate}, \\ 
        0.68 \leq \delta \leq 1.00 & \quad \text{corresponds to advanced}. 
    \end{aligned} 
\label{eq:difficulty-intervals} 
\end{equation}
These intervals provide an approximate correspondence between RAFT's continuous difficulty model and RoR's three fixed difficulty levels. They allow us to compare the scenarios presented by the adaptive and non-adaptive trainers without replacing RAFT's gradual adaptation with RoR's discrete transitions.

\subsection{Feedback}

\begin{figure}[htbp]
    \centering
    \begin{subfigure}{0.49\textwidth}
        \centering
        \includegraphics[width=\textwidth]{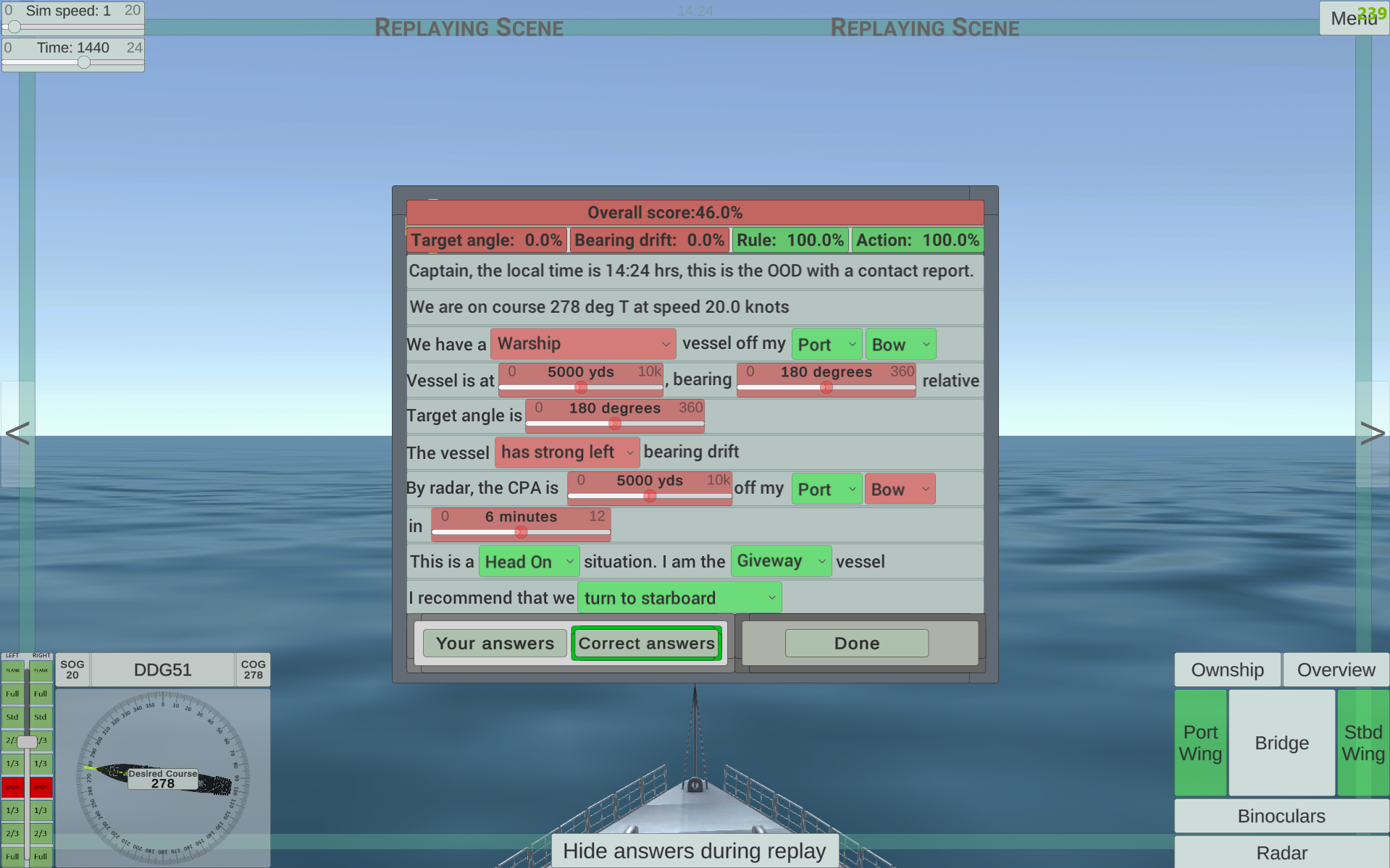}
        \caption{}
        \label{feedbackScreenshot}
    \end{subfigure}%
    \hfill
    \begin{subfigure}{0.49\textwidth}
        \centering
        \includegraphics[width=\textwidth]{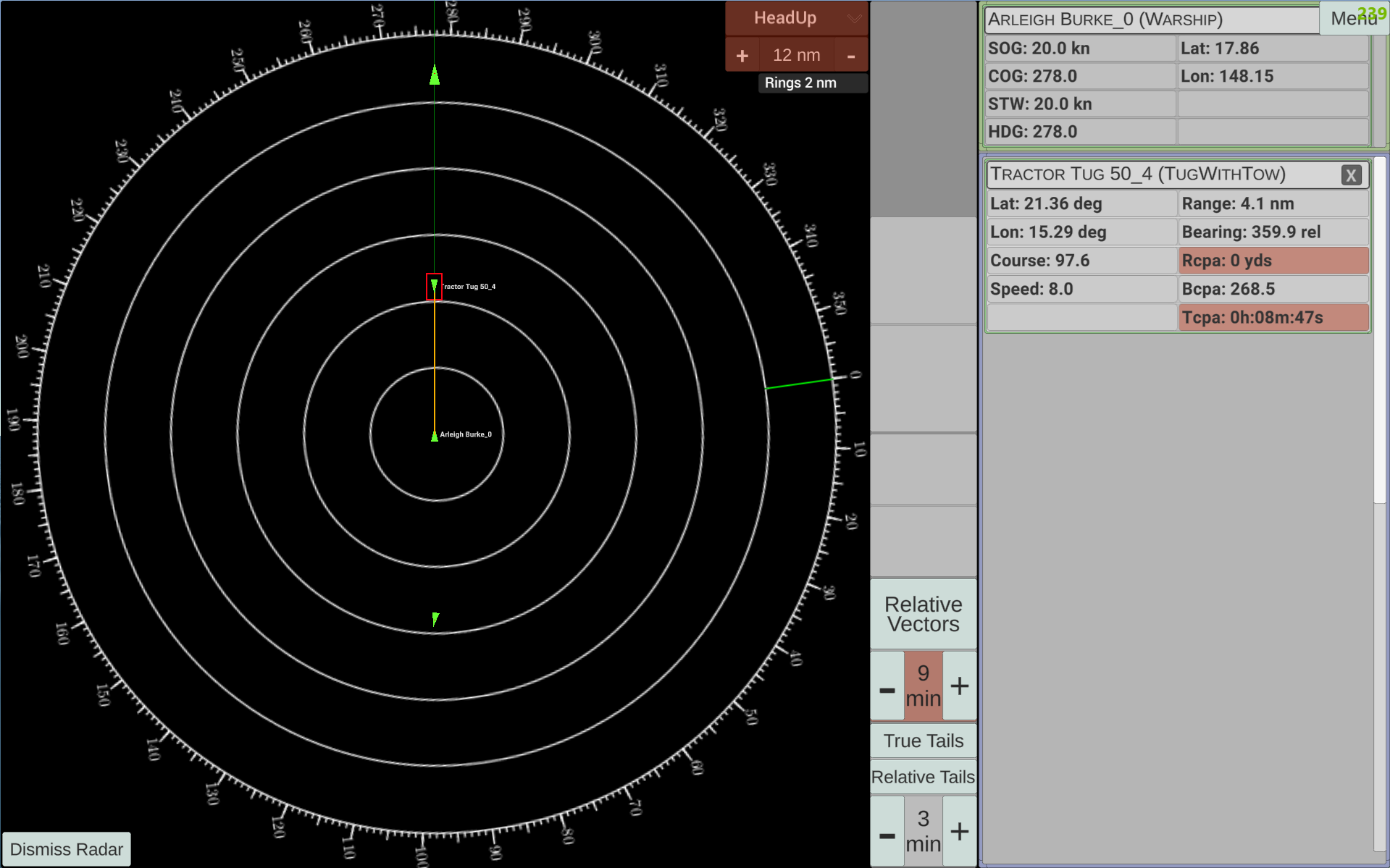}
        \caption{}
        \label{feedbackRadarScreenshot}
    \end{subfigure}
    \caption{The feedback screen (\ref{feedbackScreenshot}) shows one form of feedback provided by a graded Captain's report. Another form of feedback is provided by an annotated voyage management system display and is shown on the right (\ref{feedbackRadarScreenshot}).}
    \label{fig:FeedbackBoth}
\end{figure}

RAFT provides feedback immediately after a student submits a Captain's Report. It first displays a feedback version of the report and highlights each submitted value. Green indicates a correct response, and red indicates an incorrect response. When the report contains an incorrect response, the student can review the correct values and replay the scenario. This feedback is shown in Figure~\ref{feedbackScreenshot}. 

RAFT also presents an annotated voyage management system radar display that directs the student's attention to information relevant to the collision risk on the display. For example, Figure~\ref{feedbackRadarScreenshot} highlights the vessel that presents the greatest collision risk and identifies radar information that can help the student analyze the encounter. The student can examine the head-up display, relative-motion vectors, TCPA, and RCPA to understand why the highlighted target creates a risk of a head-on collision.
This feedback connects an incorrect answer to the visual and numerical evidence available in the scenario. Rather than only identifying an error, the training systems direct the student toward the information needed to correct it. 

\section{Experimental Design}

\begin{figure}[htbp] 
    \centering 
    \includegraphics[width=3.5in]{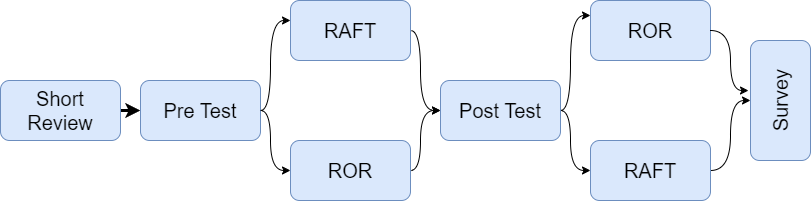} 
\caption{Experimental design. Students completed a pre-test, trained with either adaptive RAFT or non-adaptive RoR, and then completed a post-test. After the post-test, each group used the other trainer so that all students could compare the two systems in the final survey.} 
\label{experimentalDesign} 
\end{figure}

We conducted a controlled study to compare the adaptive RAFT trainer with the non-adaptive RoR trainer. We recruited 30 university students and randomly assigned 15 students to each training condition. The study lasted approximately 90 minutes and followed the sequence shown in Figure~\ref{experimentalDesign}. 

At the beginning of the study, we introduced or reviewed the relevant nautical rules of the road and demonstrated how to use the simulation trainer. All students then completed the same 10-question pre-test. Students typically required $60$ to $90$ seconds to answer each question. 

After the pre-test, students practiced with their assigned trainer. The $15$ students in the non-adaptive condition used RoR in \emph{practice quiz} mode. RoR generated scenarios at a fixed novice difficulty level, corresponding to $\delta \in [0.00, 0.33]$. Students could modify selected scenario variables, such as changing a scenario from night to day, as they practiced identifying collision risks and applying the appropriate rules of the road. The $15$ students in the adaptive condition trained with RAFT. Each student began at $\delta=0.2$, which falls near the middle of the novice difficulty range. After each scenario, RAFT updated the difficulty according to Equation~\ref{eq:difficulty-update} using the student's performance on that scenario. RAFT then used the updated value of $\delta$ to generate the next scenario. RAFT also provided immediate, context-sensitive feedback after each submitted response.

After the practice session, all students completed the same 10-question post-test. We used the pre-test and post-test results to compare learning effectiveness between the RAFT and RoR conditions. We also recorded the time students required to answer each question. We then gave students experience with the trainer that they had not used during the practice session. Students initially assigned to RAFT used RoR, and students initially assigned to RoR used RAFT. This post-test exposure did not contribute to the pre-test and post-test comparison. Instead, it allowed every student to experience both trainers before completing the final survey. The survey asked students to compare the trainers' perceived effectiveness, overall quality, engagement, feedback, and question difficulty.

\section{Results} \label{Results} 
We first examined how RAFT adapted scenario difficulty for students in the adaptive condition. We then compared scores and response times between the adaptive and non-adaptive conditions. Finally, we analyzed students' survey responses after all students had experienced both trainers. 

\begin{figure}[htbp] 
    \centering 
    \includegraphics[width=3.5in]{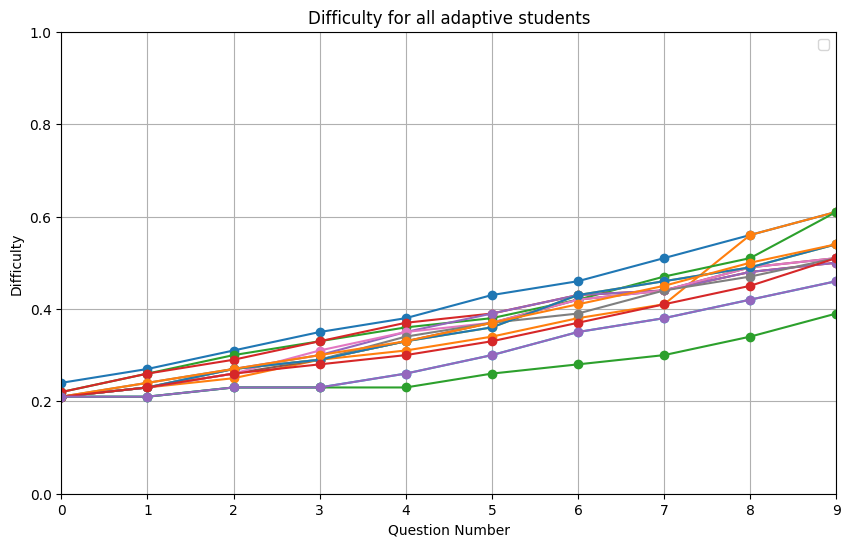} 
\caption{Difficulty trajectories for students in the RAFT condition. RAFT increased difficulty according to Equation~\ref{eq:difficulty-update}, with the rate of progression determined by each student's performance.} 
\label{difficultyChange} 
\end{figure}

Figure~\ref{difficultyChange} shows the difficulty trajectory for each student in the RAFT condition. All students began at $\delta=0.2$, and RAFT increased difficulty according to Equation~\ref{eq:difficulty-update} as students completed the training scenarios. Although difficulty generally increased across scenarios, the trajectories differed because RAFT based each update on the individual student's performance. 

By the end of training, all 15 students in the RAFT condition had progressed from the novice range to the intermediate range, $\delta \in [0.34,0.67]$. Their final difficulty values ranged from $\delta=0.4$ to $\delta=0.6$. Three students reached the highest observed final difficulty, $\delta=0.6$.

\subsection{Adaptive and Non-Adaptive Training Effectiveness} \label{sec:training-effectiveness} 
The pre-test contained 10 novice-level scenarios. The post-test also contained 10 scenarios, but it included six novice-level and four intermediate-level scenarios. To account for this difference in difficulty, we calculated a weighted score in which each intermediate-level question received twice the weight of each novice-level question. 

Table~\ref{table:scores} reports the mean normalized pre-test and post-test scores for the two training conditions. Before training, students assigned to RAFT had a mean score of $57.1\%$, while students assigned to RoR had a mean score of $62.3\%$. After training, the mean score increased to $74.3\%$ in the RAFT condition and decreased slightly to $61.4\%$ in the RoR condition. 
\begin{table}[htp]
    \centering 
    \begin{tabular}{|c|c|c|} \hline 
        \textbf{Normalized Score} & \textbf{Adaptive / RAFT} & \textbf{Non-adaptive / RoR} \\ \hline 
        Pre-test & $57.1\%$ & $62.3\%$ \\ \hline 
        Post-test & $74.3\%$ & $61.4\%$ \\ \hline 
    \end{tabular} 
\vspace*{0.2cm} 
\caption{Mean normalized pre-test and post-test scores for the adaptive RAFT and non-adaptive RoR conditions. Students in the RAFT condition achieved higher post-test scores than students in the RoR condition ($p < 0.0001$).} 
\label{table:scores} 
\end{table} 

Students in the RAFT condition achieved significantly higher post-test scores than students in the RoR condition. A two-tailed $t$-test found a significant between-condition difference ($p < 0.0001$), with a large effect size ($d > 2$). Within the RAFT condition, scores increased significantly from the pre-test to the post-test ($p < 0.0001$). We found no significant pre-test to post-test increase in the RoR condition. Instead, the mean RoR score decreased from $62.3\%$ to $61.4\%$, but this decrease was not statistically significant. 

We also compared the mean time students required to answer each question. Table~\ref{table:times} shows that the groups had similar mean response times on the pre-test. Students in the RAFT condition required an average of $91.9$ seconds per pre-test question, while students in the RoR condition required $90.0$ seconds. On the post-test, the RAFT mean decreased to $79.5$ seconds, whereas the RoR mean increased to $99.4$ seconds. 
\begin{table}[htp] 
    \centering 
    \begin{tabular}{|c|c|c|} \hline 
        \textbf{Mean Time per Question} & \textbf{Adaptive / RAFT} & \textbf{Non-adaptive / RoR} \\ \hline 
        Pre-test & $91.9$ seconds & $90.0$ seconds \\ \hline 
        Post-test & $79.5$ seconds & $99.4$ seconds \\ \hline 
    \end{tabular} 
\vspace*{0.2cm} 
\caption{Mean response times for the adaptive RAFT and non-adaptive RoR conditions. Students in the RAFT condition answered post-test questions more quickly than students in the RoR condition ($p < 0.0001$).} 
\label{table:times} 
\end{table} 

Students in the RAFT condition answered post-test questions significantly faster than students in the RoR condition. A two-tailed $t$-test found a significant between-condition difference in post-test response time ($p < 0.0001$), with a large effect size ($d > 2$). Within the RAFT condition, mean response time decreased significantly from the pre-test to the post-test. We found no significant pre-test to post-test change in the RoR condition, although the mean response time increased from $90.0$ to $99.4$ seconds. 

These results show that students who trained with RAFT achieved higher post-test scores and answered post-test questions more quickly than students who trained with RoR. The intermediate-level questions in the post-test may have contributed to these between-condition differences. We believe that intermediate level questions on the post-test contributed to these differences in scores and times between adaptive and non-adaptive students. Students in the adaptive (RAFT) condition needed less time to do significantly better on harder questions.

\subsection{Student Preferences} \label{sec:student-preferences} 
\begin{figure}[htp] 
    \centering 
    \includegraphics[height=1.5in]{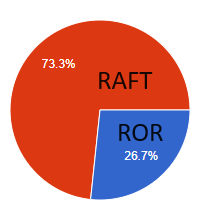} 
\caption{Students' judgments of the more effective training strategy. Twenty-two of 30 students ($73\%$) selected RAFT, and 8 students ($27\%$) selected RoR.} 
\label{overall} 
\end{figure}

After the post-test, each student used the trainer that they had not used during the initial training phase. All 30 students therefore experienced both RAFT and RoR before completing the survey. The survey measured students' perceptions of the trainers rather than their objective learning performance. We first asked students, "Overall, which strategy, Blue or Yellow, do you think was more effective for learning and assessing your knowledge of Rules of the Road?"\footnote{During the study, we used the labels Blue and Yellow to identify RAFT and RoR. These labels do not refer to the colors used in Figure~\ref{overall}.} Figure~\ref{overall} shows that 22 of the 30 students ($73\%$) selected RAFT, while 8 students ($27\%$) selected RoR. 

\begin{figure}[htbp] 
    \centering 
    \includegraphics[height=1.5in] {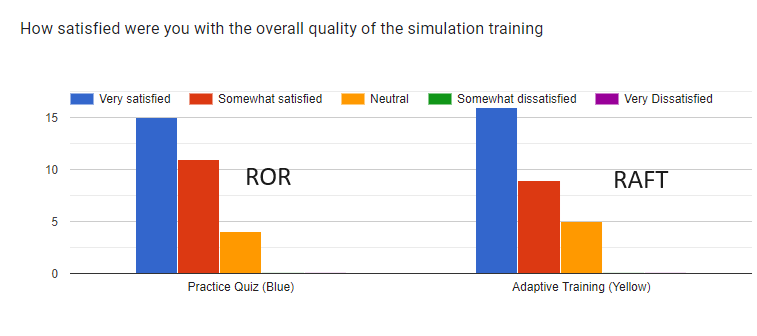} 
\caption{Students' ratings of the overall quality of the learning experience. Students generally reported satisfaction with both RAFT and RoR.} 
\label{quality} 
\end{figure}

We also asked students to rate the overall quality of the learning experience provided by each trainer. As Figure~\ref{quality} shows, students generally rated both trainers positively. No student selected "Somewhat dissatisfied" or "Very dissatisfied" for either trainer. Thus, although most students judged RAFT to be the more effective trainer, their ratings indicated general satisfaction with the overall quality of both systems. 

\begin{figure}[htbp] 
    \centering 
    \includegraphics[height=1.5in] {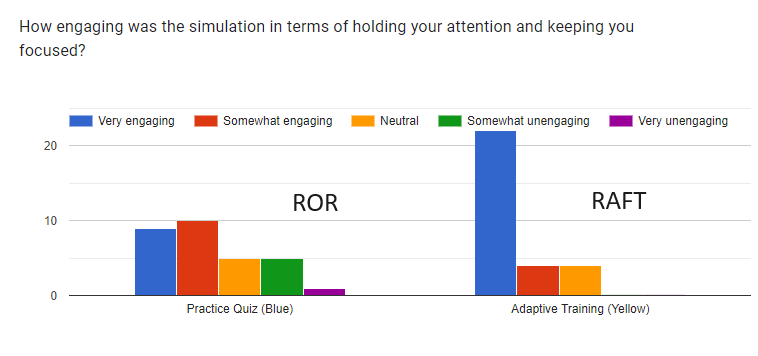} 
\caption{Students' ratings of engagement and attention. Twenty-two of 30 students rated RAFT as "Very engaging," compared with 9 of 30 students for RoR.} 
\label{engagement} 
\end{figure}

Figure~\ref{engagement} compares students' engagement ratings. Of the 30 students, 22 rated RAFT as "Very engaging," the most positive response on the scale. In comparison, 9 students gave RoR the same rating. These responses show that students perceived RAFT as more engaging than RoR. RAFT's changing scenario difficulty and more varied training experience may have contributed to this result, but the survey does not establish which system feature caused the difference. 

\begin{figure}[htp]
  \centerline{
    \includegraphics[height=1.5in]{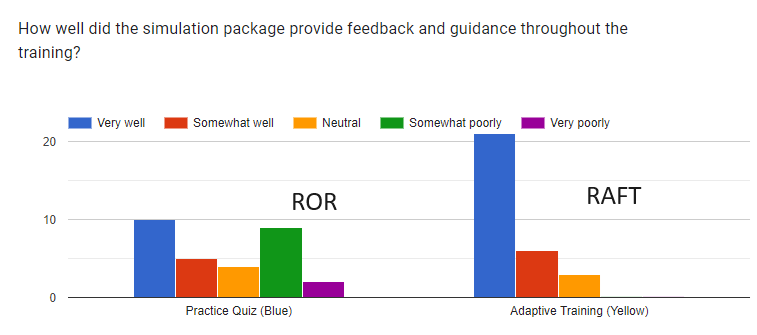}
    }
  \caption{Students' ratings of feedback and guidance. Twenty-one of 30 students selected "Very well" for RAFT, compared with 10 of 30 students for RoR.}
 \label{feedback}
\end{figure}

We next asked, "How well did the simulation package provide feedback and guidance throughout the training?" Figure~\ref{feedback} shows that 21 students selected "Very well" for RAFT, compared with 10 students for RoR. Students therefore gave RAFT's feedback more of the highest possible ratings. These survey responses are consistent with a preference for RAFT's feedback configuration.

\begin{figure}[htbp] 
    \centering 
    \includegraphics[height=1.5in] {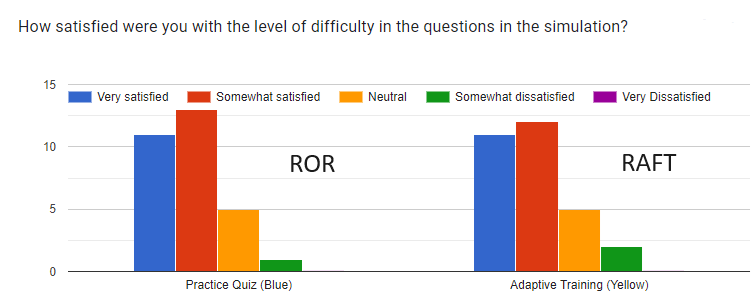} 
\caption{Students' satisfaction with question difficulty in RAFT and RoR. Students generally reported satisfaction with the difficulty of both trainers.} 
\label{qDifficulty} 
\end{figure}

Because RAFT adjusts scenario difficulty, we also asked students, "How satisfied were you with the level of difficulty in the questions in the simulation?" Figure~\ref{qDifficulty} shows similar response patterns for the two trainers. Most students reported that they were "Very satisfied" or "Somewhat satisfied" with the question difficulty in both systems, and no student reported dissatisfaction. Thus, students generally considered the difficulty levels in both trainers acceptable. 

\begin{figure}[htbp] 
    \centering 
    \includegraphics[height=1.5in] {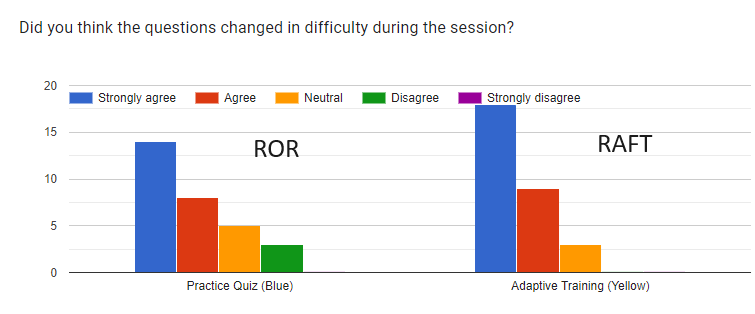} 
\caption{Students' perceptions of changes in question difficulty. Twenty-seven of 30 students agreed or strongly agreed that RAFT questions varied in difficulty, compared with 22 of 30 students for RoR.} 
\label{deltaDifficulty} 
\end{figure}

Finally, we asked students whether the questions changed in difficulty during training. As Figure~\ref{deltaDifficulty} shows, 22 of the 30 students ($73\%$) agreed or strongly agreed that the RoR questions varied in difficulty. For RAFT, 27 students ($90\%$) agreed or strongly agreed that the questions varied in difficulty. The remaining students selected neutral or negative responses. 

Overall, the survey responses show that most students judged RAFT to be more effective and gave RAFT higher ratings for engagement and feedback. Students expressed similar satisfaction with the overall quality and question difficulty of both trainers. Because students completed the survey after experiencing both systems, these responses support direct comparisons of their perceptions. They do not, however, provide an independent experimental estimate of how individual RAFT features affected learning.

\section{Conclusions and Future Work} \label{sec:conclusion} 

We developed and evaluated two simulation trainers for teaching the nautical rules of the road: the adaptive RAFT system and the non-adaptive RoR system. Both trainers used the same instructional content, scenario-generation framework, and user interface. However, they differed in how they controlled scenario difficulty and delivered feedback. RoR generated scenarios at a fixed difficulty level and provided feedback after a quiz. RAFT adjusted a continuous difficulty variable in response to student performance and provided immediate, context-sensitive feedback after each scenario. 

We conducted a study with $30$ university students to compare the two training configurations. We randomly assigned $15$ students to the RAFT condition and $15$ students to the RoR condition. All students completed a pre-test, practiced with their assigned trainer, and then completed a post-test. Students who trained with RAFT achieved significantly higher post-test scores than students who trained with RoR ($p < 0.0001$). They also answered post-test questions in significantly less time ($p < 0.0001$). Both comparisons produced large effect sizes ($d > 2$). Within the RAFT condition, scores increased and response times decreased significantly from the pre-test to the post-test. We found no significant pre-test to post-test improvement in the RoR condition. 

After the post-test, each student used the trainer that they had not used during the initial training phase. All students therefore experienced both RAFT and RoR before completing the final survey. Twenty-two of the $30$ students ($73\%$) judged RAFT to be the more effective trainer. The survey responses also favored RAFT on engagement and feedback. Twenty-two students rated RAFT as "Very engaging," compared with $9$ students who gave RoR the same rating. Similarly, 21 students rated RAFT's feedback and guidance "Very well," compared with $10$ students for RoR. Students nevertheless reported general satisfaction with the overall quality and question difficulty of both trainers. 

Together, the performance and survey results provide evidence that RAFT adaptive training supported better learning outcomes and a more engaging training experience than the non-adaptive RoR in this study. However, the sample consisted of $30$ university students, so additional studies must determine whether the findings generalize to other populations and operational settings. 

Many non-adaptive training systems already represent task difficulty through variables that an instructor or learner selects. RAFT illustrates one way to convert such a fixed-level system into a performance-based system. Instead of selecting among a few discrete difficulty levels, RAFT maintains a continuous difficulty value, updates that value from student performance, and maps it to variables that control the next scenario. This approach allows the trainer to increase scenario difficulty gradually as a student progresses. 

Future work will extend RAFT in several directions. First, we will investigate adaptation methods that can both increase and decrease difficulty in response to student performance. Second, we will study the separate effects of difficulty adaptation and immediate feedback through experimental conditions that vary these features independently. Third, we will move beyond a single difficulty value by adapting multiple scenario variables independently. This approach may allow RAFT to identify specific weaknesses and generate scenarios that target them. Finally, we will investigate whether data from previous students can improve the initial training configuration for new students or cohorts.

\subsection*{Acknowledgments}

This work was supported by grant number N00014-22-1-2122 from the Office of Naval Research. Any opinions, findings, and conclusions or recommendations expressed in this material are those of the author(s) and do not necessarily reflect the views of the Office of Naval Research.

\bibliographystyle{ieeetr} 
\bibliography{ref}

\end{document}